\documentstyle[seceq,preprint,epsbox]{jpsj}
\def\runtitle{Role of Vertex Correction and Dimensionality on Superconductivity Induced by On-site Coulomb Repulsion}

\def\runauthor{Hirono {\sc Fukazawa}, Hiroaki {\sc Ikeda} and Kosaku {\sc Yamada}}
\title
{Role of Vertex Correction and Dimensionality on Superconductivity Induced by On-site Coulomb Repulsion}

\author
{Hirono {\sc Fukazawa} \footnote{E-mail: hirono@scphys.kyoto-u.ac.jp},
 Hiroaki {\sc Ikeda} and Kosaku {\sc Yamada}}

\inst
{Department of Physics, Kyoto University, Kyoto 606-8502.}

\recdate
{\today}
\abst
{
We evaluate the dominant superconducting pairing symmetry in some 3-dimensional cubic lattice structures within the third-order perturbation theory with respect to the on-site coulomb repulsion. We clarify whether the vertex correction term, which has a critical contribution to the p-wave state in the 2-dimensional systems, is also important in the 3-dimensional system. We finally investigate the interplay between dimensionality and the vertex correction to grasp common features independent of the lattice structures in the mechanism of the superconductivity.
}
\kword
{
Hubbard model, vertex correction, superconductivity, dimensionality, lattice structure, pairing symmetry, strong coupling}
\begin{document}
\sloppy
\maketitle

\section{Introduction}

In the strongly correlated fermion systems, the origin of superconductivity has been investigated with the special interest. In quasi-2-dimensional systems, such as the high-$T_{\rm c}$ cupretes$^{1-4)}$, the $d$-wave superconductivity is considered to originate from the antiferromagnetic spin fluctuation. Such a spin fluctuation is a candidate for the mechanism of the superconductivity in the strongly correlated systems. Recently, on the basis of the spin fluctuation mechanism, the possibility of the $p-$ and $d-$wave pairing states has been investigated in a variety of materials$^{5-9)}$. For example, Arita $et~al.$\cite{rf:5} studied the pairing instabilities mediated by the spin fluctuation with the fluctuation exchange (FLEX) approximation, in the 2- and 3-dimensional Hubbard model. In their results, $d$-wave pairing is easily induced by the antiferromagnetic spin fluctuation in the high density. The $p$-wave is dominant due to the ferromagnetic spin fluctuation, in very low density. As their conclusion, the $p$-wave instability mediated by the ferromagnetic spin fluctuations is weaker than the $d$-wave instability mediated by the antiferromagnetic spin fluctuations in both 2- and 3-dimensional systems. Actually, the simple spin fluctuation theory does not lead to the $p$-wave triplet state in Sr$_2$RuO$_4$\cite{rf:10}. This indicates that the simple spin fluctuation mechanism is not the unified theory, but a limited one in the strongly correlated superconductors. On the other hand, the third-order perturbation theory (TOPT) with respect to the on-site Coulomb repulsion $U$ can most naturally explain the triplet state in Sr$_2$RuO$_4$\cite{rf:11}. In this case, the $p-$wave triplet state is promoted by the wave-number dependence of the third-order vertex correction rather than the spin fluctuation process.

Motivated by this fact, in the previous paper\cite{rf:12}, we have investigated in detail the effect of the vertex correction in the 2-dimensional square lattice. Then, we have determined the pairing symmetry in the phase diagram for electron density and next-nearest-neighborer hopping parameter. We have shown that the triplet state induced by the vertex correction term is dominant in the wide range of electron density. Thus, we have indicated that the vertex correction term has a critical contribution to the triplet state in the superconductivity induced by the on-site Coulomb repulsion in the 2-dimensional systems. This conclusion continuously presents a question whether the vertex correction is also important in the 3-dimensional systems. We in this paper evaluate the dominant pairing symmetry in some cubic lattice structures, and investigate the interplay between dimensionality and the vertex correction to grasp common features independent of lattice structures in the mechanism of the superconductivity.

Before starting our study, let us discuss possibility of the magnetic instability. The strict ground state, actually, is determined by comparing total energies in the superconducting and the magnetic states. However, since our aim is not to obtain the complete phase diagram, we hereafter ignore the possibility of the magnetic phase transition.

This article is organized as follows. In $\S2$, we explain the formalism of the calculation. In $\S3$, we describe the results of the calculation. We show the phase diagram and discuss the role of the vertex correction in detail. In $\S4$, we summarize the interplay between dimensionality and the vertex correction to grasp common features independent of the lattice structures in the mechanism of the superconductivity, comparing with that of the simple spin fluctuation$^{5-9)}$. Finally, we mention relation of our results to the actual materials.

\section{Formulation}

The on-site repulsive Hubbard Hamiltonian in cubic lattice structures is given by
\begin{equation}
{\cal H}=-t_1\sum_{<i,j>,\sigma}c^{\dag}_{i,\sigma}c_{j,\sigma}+t_2\sum_{<i,k>,\sigma}c^{\dag}_{i,\sigma}c_{k,\sigma}+U \sum_{i}n_{i,\uparrow}n_{i,\downarrow},
\end{equation}
where $\sigma$ is the spin index, $<i,j>$ indicates taking summation over the nearest-neighbor sites and $<i,k>$ over the next-nearest-neighbor sites. We obtain energy dispersions from the non-interacting part in eq. (2.1);
\begin{eqnarray*}
{\rm E}^{SC}_k=-2t_1(\mbox{cos}k_x+\mbox{cos}k_y+\mbox{cos}k_z)+4t_2\mbox{cos}k_x\mbox{cos}k_y\mbox{cos}k_z,
\end{eqnarray*}
\begin{eqnarray*}
\hspace{-4.4cm}{\rm E}^{BCC}_k=-8t_1(\mbox{cos}k_x\mbox{cos}k_y\mbox{cos}k_z)+2t_2(\mbox{cos}(2k_x)+\mbox{cos}(2k_y)+\mbox{cos}(2k_z)),
\end{eqnarray*}
\begin{equation}
{\rm E}^{FCC}_k=-4t_1(\mbox{cos}k_x\mbox{cos}k_y+\mbox{cos}k_y\mbox{cos}k_z+\mbox{cos}k_z\mbox{cos}k_x)+2t_2(\mbox{cos}(2k_x)+\mbox{cos}(2k_y)+\mbox{cos}(2k_z)),
\end{equation}
for the simple cubic (SC), body-centered cubic (BCC) and face-centered cubic (FCC), respectively. To compare the result of the cubic lattice with our previous result of the 2-dimensional square lattice structure (SQ), we mention the case of the SQ lattice in the following. The dispersion of the SQ lattice is given by ${\rm E}^{SQ}_k=-2t_1(\mbox{cos}k_x+\mbox{cos}k_y)+4t_2\mbox{cos}k_x\mbox{cos}k_y$. We take $t_1$=1.0 here. By using the above dispersion relations, we obtain the bare Green's function, $G_0(k,\epsilon_n)=\frac{1}{{\rm i}\epsilon_n-(E_k-\mu)}$, where $\epsilon_n=\pi T(2n+1)$ is the Matsubara frequency and $\mu$ is the chemical potential. E$_k$ is the dispersion relation corresponding to each lattice structure. The particle number density $n$ per spin is given by $n=\frac{T}{N}\Sigma_{k,n}G_0(k,\epsilon_n)$.

An effective pairing interaction $V_{\rm TOPT}$ is given by the perturbation expansion up to the third-order term with respect to the on-site Coulomb interaction$^{11-15)}$. The origin of the superconductivity is investigated by all effective pairing interaction of $V_{\rm TOPT}$. However, we dare to divide it into two parts for convenience, to analysis the role of $V_{\rm TOPT}$ in detail;
\begin{equation}
V_{\rm TOPT}(q,k)=V_{\rm RPA}(q,k)+V_{\rm Vertex}(q,k).
\end{equation}
The RPA-like term $V_{\rm RPA}$ respects the term included by the Random Phase Approximation (RPA) and $V_{\rm Vertex}$ is the vertex correction. The RPA-like term reflects the nature of the simple spin fluctuations. The vertex correction term originates from the electron correlation other than the spin fluctuations. For the singlet pairing, the RPA-part and the vertex correction part are respectively given by
\begin{equation}
{\it V}_{\rm RPA}^{\rm Singlet}(q,k)=U+U^2\chi_0(q-k)+2U^3\chi_0^2(q-k),
\end{equation}
\vspace{-0.5cm}
\begin{eqnarray*}
V_{\rm Vertex}^{\rm Singlet}(q,k)={\frac{T}{N}}[ U^3[\sum_{k'}G_0(q+k'-k)\times(\chi_0(q+k')-\phi_0(q+k'))G_0(k')]
\end{eqnarray*}
\begin{equation}
\hspace{4.7cm}+ U^3[\sum_{k'}G_0(-q+k'-k)\times(\chi_0(-q+k')-\phi_0(-q+k'))G_0(k')]],
\end{equation}
where $k$ indicates $k \equiv$ (\mbox{\boldmath $k$}, $\omega_n$). The bare susceptibility $\chi_0(q)$  and $\phi_0(q)$ are defined respectively by 
\begin{equation}
\chi_0(q)=-{\frac{T}{N}}\sum_{k}G_0(k)G_0(q + k),
\end{equation}
\begin{equation}
\phi_0(q)=-{\frac{T}{N}}\sum_{k}G_0(k)G_0(q - k).
\end{equation}
For the singlet pairing, the Coulomb interaction $U$ connects only the electron lines which have the opposite spins. The diagrams for the pairing interaction are shown in Fig. 1. The two external lines have the opposite spins.

For the triplet pairing, the RPA-like term and the vertex correction correspond to the second-order and the third-order terms, respectively.
\begin{equation}
V_{\rm RPA}^{\rm Triplet}(q,k)=-U^2\chi_0(q-k),
\end{equation}
\vspace{-0.5cm}
\begin{eqnarray*}
V_{\rm Vertex}^{\rm Triplet}(q,k)={\frac{T}{N}} [U^3[\sum_{k'}G_0(q-k+k')\times(\chi_0(q+k')+\phi_0(q+k'))G_0(k')]
\end{eqnarray*}
\begin{equation}
\hspace{4.7cm}+ U^3[\sum_{k'}G_0(-q-k+k')\times(\chi_0(-q+k')+\phi_0(-q+k'))G_0(k')]].
\end{equation}
The diagrams giving the effective interaction for the triplet pairing are shown depend on the direction of $D$-vector. When the two external lines have the parallel spins, the typical diagrams are illustrated in Fig. 2. Here, we add the proviso of the diagram in Fig. 1 and Fig. 2. At first sight, it seems that the diagram of triplet pairing in Fig. 2 is different from that of singlet pairing in Fig. 1. However, the diagram in Fig. 2 is identical with that of Fig. 1.; The first-order diagram of ${\it V}_{\rm RPA}^{\rm Singlet}$ in Fig. 1 vanishes due to the triplet pairing symmetry. The second and third-order diagram of ${\it V}_{\rm RPA}^{\rm Singlet}$ cancel out each other in the triplet pairing case. Consequently, the residual singlet diagram in Fig. 1 agrees with the triplet diagram in Fig. 2.

An anomalous self-energy is written by the effective interaction $V(q,k)$ and an anomalous Green function $F(k)$, as $\Sigma_A(q)=-\frac{T}{N}\Sigma_{k}F(k)V(q,k)$. At the transition temperature $T_{\rm c}$, the value of the anomalous self-energy $\Sigma_A$ is small and we linearize the \'Eliashberg equation with respect to $F$ and $\Sigma_A$, as $F(k)^{\dagger}=|G_0(k)|^2\Sigma_A(k)^{\dagger}$. From these formulae, we obtain the following equation for the anomalous self-energy;
\begin{equation} 
\lambda\Sigma_A^{\dagger}(q)=-\frac{T}{N}\sum_{k}V(q, k)|G_0(k)|^2\Sigma_A^{\dagger}(k).
\end{equation}
This equation is the linearized \'Eliashberg equation, which is an eigenequation with an eigenvalue $\lambda$ and an eigenvector $\Sigma_A^{\dagger}$. $V(q,k)$ is given by (2.3), (2.4), (2,5), (2.8) and (2.9). We solve the linearized \'Eliashberg equation on the assumption that $\Sigma_A^{\dagger}$ has the pairing symmetry shown in Table I. 
\begin{fulltable}
\caption{Pairing symmetry}
\begin{fulltabular}{@{\hspace{\tabcolsep}\extracolsep{\fill}}llc} \hline
{\bf \large pairing symmetry} & 
{\bf \large function of symmetry} & 
{\bf \large degree of degeneracy}
\\ \hline
{\large$ d_{x^2-y^2}$} & {\large $\mbox{cos}(k_x)-\mbox{cos}(k_y)$} & {\large 2}\\

{\large$ d_{xy}$} & {\large $\mbox{sin}(k_x)\mbox{sin}(k_y)$} & {\large 3}\\
{\large$ p_{x}$} & {\large$ \mbox{sin}(k_x)$} & {\large 3}\\ 
{\large$ f_{xyz}$} & {\large$ \mbox{sin}(k_x)\mbox{sin}(k_y)\mbox{sin}(k_z)$} & {\large 1}\\ 
{\large$ f_{x(3z^2-r^2)}$} & {\large$ \mbox{sin}(k_x)(2 \mbox{cos}(k_z)-\mbox{cos}(k_x)-\mbox{cos}(k_y))$} & {\large 3} \\ 
{\large$ f_{z(x^2-y^2)}$} & {\large$ \mbox{sin}(k_z)(\mbox{cos}(k_x)-\mbox{cos}(k_y))$} & {\large 3} \\ 
{\large$ g_{xy(x^2-y^2)}$} & {\large$ \mbox{sin}(k_x)\mbox{sin}(k_y)(\mbox{cos}(k_x)-\mbox{cos}(k_y))$} & {\large 3}
\\ \hline
\end{fulltabular}
\end{fulltable}
We take up all $p$-, $d$- and $f$-wave pairing symmetries. The $g$-wave is high energy, and the possibility of the $g$-wave pairing is not stronger than that of $p$-, $d$- and $f$-wave pairing in the actual superconductivity. Therefore, we investigate the $g$-wave pairing symmetry in detail. We calculate the $g_{xy(x^2-y^2)}$-wave pairing as the typical example of the $g$-wave pairing symmetry, to obtain the full superconducting phase diagrams. The pairing symmetries are degenerate due to the space symmetry in the cubic lattice structure. We show the degree of degeneracy in the cubic lattice structure in Table. I. Here, we mention the possibility that the degenerate pairing symmetries are mixed below $T_{\rm c}$. 

The most dominant pairing symmetry has the largest value of the eigenvalues among different symmetries. When the eigenvalue $\lambda$ reaches unity, the superconducting state is realized. We solve the equation to obtain the dominant state and determine the superconducting transition temperature $T_{\rm c}$. The transition temperature $T_{\rm c}$ is higher than the actual transition temperature, because the effect of the normal self-energy is not considered in this calculation.

\section{Calculated Results}
We divide the first-Brillouin-zone into 64$\times $64$\times $64 momentum meshes and take $N_{\rm f}$ = 512 for Matsubara frequency. The bandwidth $W$ is a necessary region of Matubara frequency $\omega_n$ for reliable calculations over the low density. The region is covered with the condition; $|W|< \pi T N_{\rm f}$. To satisfy the condition, we calculate in the region with $T>$0.004 for the SC lattice structure, $T>$0.006 for the BCC lattice structure and $T>$0.008 for the FCC lattice structure. In addition to this, the condition for $U$ is $U<W$.

We obtain the superconducting phase diagram for the electron density $n$ and the next-nearest hopping integral $t_2$. Moreover, we clarify the mechanism of the superconductivity from the eigenvalue of \'Eliashberg equation. It depends on the Fermi surface, the bare susceptibility $\chi_0$ and the density of state $\rho$. Finally, we show the behavior of $T_{\rm c}$ for various values of $U$.

\subsection{Overview for Superconducting Phase Diagram}
In Fig. 3, we show the superconducting phase diagrams determined by the TOPT in the plane of the hopping integral $t_2$ and the density $n$. We figure the most dominant pairing state which has the largest value of eigenvalues $\lambda$. The lattice structures are 2-dimensional SQ and 3-dimensional SC, BCC and FCC lattice structures. We can find the following important factors in common with the various lattice structures for the SQ, SC, BCC and FCC lattice structures.

Near the half-filling $(n=0.5)$, the singlet state becomes dominant by the RPA-like term reflecting the simple spin fluctuation process. In the SC lattice, the singlet pairing also exists for $t_2$=-0.5 - 0.2 and the intermediate density.

For the triplet state, the vertex correction plays the vital role for the triplet pairing in the intermediate density ($0.075<n<0.3$). On the other hand, in the very low density ($n<$0.075), the triplet pairing state is dominant owing to the simple ferromagnetic spin fluctuations, which follows the paramagnon theory$^{16,17)}$ The triplet superconductivity can realize in the wide region of the density by adding the vertex correction, in contrast to the results$^{5-9)}$ on the basis of only the ferromagnetic fluctuations. All phase diagrams in Fig. 3 suggest that the intermediate density is necessary for the triplet pairing owing to the vertex correction, in common with the 2-dimensional SQ, 3-dimensional SC, BCC and FCC lattice structures.

Both triplet and singlet superconducting states originate from the general effective interaction including both the RPA-like term and the vertex correction. Therefore, the general effective interaction obtained by our perturbation theory gives the unified understanding on the mechanism of various superconducting states. In the following subsections, we detail the results for the pairing symmetry in the SC, BCC and FCC lattice structures.

\subsection{Simple Cubic Lattice}
\hspace{-0.3cm}{\bf $\diamondsuit$ Superconducting phase diagram}

Fig. 3(b) is the superconducting phase diagram of the SC lattice structure. Near the half-filling, the $d_{x^2-y^2}$-wave pairing state is dominant. Moreover, the $d_{x^2-y^2}$- and $d_{xy}$-wave pairings also exist for $t_2$=-0.5 - 0.2 and the large $t_2$, respectively, from intermediate to high density . 

The $p$ and $f_{xyz}$-wave pairing states are realized in the intermediate density ($0.075<n<0.3$). The triplet state is promoted by the wave number dependence of the vertex correction. In particular, the $f_{xyz}$-wave pairing is dominant for the large $t_2$. Around the $f_{xyz}$-pairing region, the $p$-wave pairing is dominant owing to the RPA-like term reflecting the ferromagnetic spin fluctuation. 
\\
\\
{\bf $\diamondsuit$ Behaviors of eigenvalue and bare susceptibility}

In Fig. 4, we show the $n$- and $t_2$-dependence of the eigenvalues $\lambda$ obtained by TOPT, the vertex correction term and the RPA-like term, respectively, for the $p$- and $d_{x^2-y^2}$-wave pairing states. In Fig. 5, we show the bare susceptibility $\chi_0(q,\omega_n=0)$ and the Fermi surface for the various density $n$. From the behaviors of $\lambda$ and $\chi_0$, we clarify each effect of the RPA-like term $V_{\rm RPA}$ and the vertex correction $V_{\rm Vertex}$.

The behaviors of $\lambda$ in Fig. 4(a) shows that the $d_{x^2-y^2}$-wave pairing is induced by the RPA-like term near the half-filling. The vertex correction suppresses the $d_{x^2-y^2}$-wave pairing. For the $d_{x^2-y^2}$-wave pairing, $\chi_0$ in Fig. 5(a) shows the spin fluctuations corresponding to a peak of $\chi_0$ at the momentum $q=(\pi,\pi,\pi)$. The antiferromagnetic spin fluctuation originates from a nesting feature of the Fermi surface and is reflected by the RPA-like term. Thus, the RPA-like term promotes the attractive force for the $d_{x^2-y^2}$-wave pairing state. 

For the $d_{xy}$-wave pairing, the spin fluctuations at $q=(0,\pi,0),~(\pi,0,0),~(0,0,\pi)$ are reflected by the RPA-like term in Fig. 5(b). Thus, the $d_{xy}$-wave pairing is dominant for the large next-nearest hopping integral $t_2$, which gives a strong nesting feature. The vertex correction also suppresses the $d_{xy}$-wave pairing.
 
In Fig. 4(b), 4(c) and 4(d), we show the behaviors of $\lambda$ for the $p$- and $f_{xyz}$-wave pairing states. When the solution of the \'Eliashberg equation does not converge, we set $\lambda$ zero. For the intermediate density, the $p$- or $f_{xyz}$-wave pairing state has the largest eigenvalue owing to the vertex correction.

The superconducting mechanism due to the vertex correction is explained in the following. The vertex correction has a unique wave number dependence. The wave number dependence of the vertex correction is not owing to the simple spin fluctuation process, because any character of the spin fluctuation does not clearly appear in Fig. 5(c) and 5(d). The characteristic wave number dependence of the vertex correction is induced by the general particle-particle iteration due to the Coulomb interaction. The wave number dependence induces the attractive force between the electrons near the Fermi surface. The RPA-like term suppresses the eigenvalue for the $p$- and $f$-wave pairing states. 
\\
\\
{\bf $\diamondsuit$ Behaviors of eigenvalue and density of states}

In Fig. 4(b), $\lambda$ for the $p$-wave pairing is large from $n$=0.175 to 0.275. This is because the density of states $\rho$ is large at Fermi level from $n$=0.175 to 0.275 in Fig. 6(a). In Fig. 4(c), $\lambda$ for the $p$-wave pairing is small as $t_2$ is small. The decrease originates from the result that $\rho$ at Fermi level decreases as $t_2$ becomes small, which is shown in Fig. 6(b).

\subsection{Body-centered Cubic Lattice}
\hspace{-0.3cm}{\bf $\diamondsuit$ Superconducting phase diagram}

Fig. 3(c) is the superconducting phase diagram of the BCC lattice structure. In the high density near the half-filling, the $d_{xy}$-wave pairing is induced by the RPA-like term which reflects the simple spin fluctuations. In the intermediate density ($0.075<n<0.3$), the $p$-wave pairing is promoted by the wave number dependence of the vertex correction. The $f$- and $g$-wave pairing states do not appear in this case.
\\
\\
{\bf $\diamondsuit$ Behaviors of eigenvalue and bare susceptibility}

In Fig. 7(a) and 7(b), we show the $n$- and $t_2$-dependence of the eigenvalue $\lambda$ obtained for the $p$-wave pairing state. The behavior of $\lambda$ is similar to the case of the SC lattice structure. The $p$-wave pairing state is dominant owing to the vertex correction in the intermediate density ($0.075<n<0.3$). The RPA-like term suppresses the $p$-wave pairing. On the other hand, the $d_{xy}$-wave is dominant owing to the RPA-like term near the half-filling, while the vertex correction suppresses the $d_{xy}$-wave pairing. This mechanism for the $p$- and $d$-wave pairing states are similar to that in the case of the SC lattice.

In Fig. 8, we show the bare susceptibility $\chi_0(q,\omega_n=0)$ and the Fermi surfaces for $n$=0.45 and 0.15 at $t_2$=0.1, respectively. In Fig. 8(a), the antiferromagnetic spin fluctuations at $q=(\pi, \pi, \pi),~(\pi, 0, 0)$ and $(0, \pi, 0)$ encourage the $d_{xy}$-wave pairing state. The antiferromagnetic spin fluctuations are induced by the nesting feature of the Fermi surface, such as the $d_{xy}$-wave pairing case of the SC lattice. In Fig. 8(b), $\chi_0(q, \omega_n=0)$ does not have any character of the spin fluctuation but the vertex correction induces the $p$-wave pairing. 
\\
\\
{\bf $\diamondsuit$ Behaviors of eigenvalue and density of states}

We show the $n$- and $t_2$-dependence of the density of states $\rho$ in Fig. 9. $\lambda$ is large from $n$=0.4 to 0.475 in Fig. 7(a), because the density of states $\rho$ in Fig. 9(a) is large at the Fermi level. The $\lambda$ in Fig. 7(b) becomes small as $t_2$ is small. This is because that the density of states at Fermi level decreases as $t_2$ becomes small, which is shown in Fig. 9(b). 

\subsection{Face-centered Cubic Lattice}
\hspace{-0.3cm}{\bf $\diamondsuit$ Superconducting phase diagram}

In Fig. 3(d), we show the superconducting phase diagram for the FCC lattice structure. Near the half-filling, the $d_{xy}$- or $g$-wave pairing state is dominant by the RPA-like term. The $p$-wave pairing state is promoted by the vertex correction in the intermediate density ($0.075<n<0.3$). In contrast with this, the $p$-wave pairing is induced by the RPA-like term for the low hole density ($n>0.7$).
\\
\\
{\bf $\diamondsuit$ Behaviors of eigenvalue and bare susceptibility}

In Fig. 10, we show the $n$- and $t_2$-dependence of the eigenvalue $\lambda$ for the $p$-wave pairing state. In the intermediate density ($0.075<n<0.3$), the $p$-wave pairing state is dominant. The RPA-like term suppresses the $p$-wave pairing. On the other hand, the RPA-like term encourages the triplet pairing in the low hole or low electron density ($n>0.7$ or $n<0.075$). Thus, the ferromagnetic fluctuations play the important role in the triplet superconductivity. The vertex correction suppresses the $p$-wave pairing. 

In Fig. 11, the bare susceptibility $\chi_0(q,\omega_n=0)$ and the Fermi surface are shown. Near the half-filling, there are the spin fluctuations corresponding to a peak of $\chi_0$ at a certain momentum vector in Fig. 11(a), 11(b) and 11(c). The spin fluctuations are reflected by the RPA-like term and the term gives the advantage for the $d_{x^2-y^2}$-, $d_{xy}$- and $g$-wave pairing states. The vertex correction suppress the singlet pairing states. The spin fluctuations are induced by a feature of the Fermi surface. The spin fluctuations at $q=(\pi/2, \pi/2, k_{\rm z})$ encourage the $d_{x^2-y^2}$-wave pairing state in Fig. 11(a). On the other hand, the spin fluctuations at $q=(0, 0, \pi),~(0, \pi, 0),~(\pi, 0, 0)$ encourage the $d_{xy}$-wave pairing state in Fig. 11(b).

For the intermediate density ($0.075<n<0.3$) in Fig. 11(d), the feature of a certain spin fluctuation does not clearly appear. The wave number dependence of the vertex correction induce the $p$-wave pairing. 

For the low hole or low electron density, $\chi_0(q, \omega_n=0)$ has the ferromagnetic spin fluctuation, which exhibits a peak of $\chi_0$ near $q=(0, 0, 0)$ in Fig. 11(e). The ferromagnetic spin fluctuation gives the advantage of the $p$-wave pairing state. The situation of the low hole density is connected with the characteristic density of states in the FCC lattice, which also induces the ferromagnetism. The detail of explanation is the following.
\\
\\
{\bf $\diamondsuit$ Behaviors of eigenvalue and density of states}

The density of states $\rho$ is large at the Fermi level as shown in Fig. 12(a) in the low hole density ($n=0.84$). This is the particular nature of the density of states in the FCC lattice structure. On the basis of the particular band structure, the ferromagnetism of Ni is explained by Kanamori$^{18}$. In the low hole density, the $p$-wave pairing is dominant owing to the ferromagnetic fluctuations, which agrees with the Arita's results$^{5)}$.

However, the triplet pairing induced by the ferromagnetic spin fluctuation does not have the large eigenvalue in the \'Eliashberg equation. This fact indicates that it is not easy for the simple ferromagnetic spin fluctuation to introduce the triplet pairing state. Moreover, the eigenvalue is dominant due to the ferromagnetic spin fluctuation and is suppressed by the vertex correction, as the electron correlation becomes strong. Therefore, the superconducting transition temperature is very low for the realistic superconductor to exist owing to the ferromagnetic spin fluctuation.  

In Fig. 10(c), $\lambda$ also decreases as $t_2$ becomes small. The decrease originates from the fact that $\rho$ at Fermi level decreases as $t_2$ becomes small as shown in Fig. 12(b).

\subsection{Dependence of Eigenvalue on $U$ and effect of vertex correction}

In this subsection, we certify that the triplet pairing induced by the vertex correction exists in the wide region of $U$. 

We show the $U$-dependence of the eigenvalue $\lambda$ obtained on the basis of TOPT in Fig. 13. We focus the density $n$ at the crosspoint between the $p$-wave and the $d$-wave pairing. The $p$-wave pairing is dominated by the vertex correction. The crosspoint shifts to small $n$ as $U$ becomes small from 7.0 to 2.0. The bandwidth $W$ in the SC lattice structure equals 12. Thus, the density region of the $p$-wave pairing state becomes narrow, while the density region of the $d$-wave spreads. The density $n$ at the crosspoint changes from 0.22 to 0.32 in Fig. 13. The change is not so large, and the phase diagram in Fig. 3(b) does not change so much for the value of $U$, as seen in our previous result\cite{rf:12} of the 2-dimensional SQ lattice. Therefore, the $p$-wave due to the vertex correction exists in the intermediate density in the wide region of $U$. For the small $U/W$ such as $U=2.0$ in Fig. 13(a), the effect of the higher-order interactions is weaker than that of the second- and third-orders. (For $U<2.0$, we do not calculate $\lambda$, because $\lambda$ is too small to obtain a reliable calculation.) Even if we take into account of the effect of the higher-order interactions, the $p$-wave induced by the vertex correction seems to be dominant in the intermediate density. This fact gives reliability to the result of TOPT.

\subsection{Transition Temperature $T_{\rm c}$ and dimensionality}
We discuss the effect of the dimensionality on $T_{c}$. In Fig. 14, we show the $n$-dependence of the superconducting transition temperature $T_{\rm c}$ calculated by TOPT for the $p$- and $d_{x^2-y^2}$-wave pairing states. The parameters are $t_2$=0.1 and $U$=10.0. The unit of energy is the hopping transfer $t_1$. $T_{\rm c}$ in the 3-dimensional SC lattice is one-order lower than $T_{\rm c}$ in the 2-dimensional SQ lattice structure\cite{rf: 12}. Therefore, both singlet and triplet pairing states are in favor of the 2-dimensional system such as the layered system rather than the 3-dimensional system.

\section{Summary and Conclusion}
We have studied the superconductivity in the Hubbard model by the third-order perturbation theory. We conclude by pointing out the main factors in common with the 2- and 3-dimensional various lattice structures; the SQ, SC, BCC and FCC lattice structures.
\\
\\
{\bf (I) \it Study on the basis of effective interaction including vertex correction}
\\
{\bf $\diamondsuit$ Singlet pairing}

For the singlet pairing, the spin fluctuation is the important factor. Near the half-filling or the intermediate density, the singlet superconductivity is realized by the spin fluctuation, such as the antiferromagnetic spin fluctuation. The vertex correction suppresses the singlet superconductivity induced by the spin fluctuation.

In the 3-dimensional system, the effect of the magnetic order is strong near the half-filling. Therefore, it seems to be difficult to realize the unconventional singlet superconductivity. On the other hand, in the heavy fermion system, the magnetic instability is usually suppressed by the Kondo effect\cite{rf: 19}. In this case, the singlet pairing is dominant around the half-filling case in the superconducting phase diagram of the SC lattice. The cubic heavy fermion superconductor CeIn$_3$ may be a good example.
\\
\\
{\bf $\diamondsuit$ Triplet pairing}

By taking account of the vertex correction, the triplet pairing can be realized in the wide region of the electron density. The triplet pairing is induced by the vertex correction in the intermediate density.

On the other hand, the $p$-wave pairing is also induced by the ferromagnetic spin fluctuation process for the low hole or low electron density. In particular, the case of the low hole density in the FCC lattice has a similar electronic structure to that of the ferromagnetism of Ni and nearly ferromagnetic Pd. The density of states in the FCC lattice has the narrow and high density peak at the Fermi level for the low hole density. The narrow and high density of states gives rise to the ferromagnetic spin fluctuation, which induces the triplet pairing state. This case of FCC lattice is the particular case in contrast with the SQ, SC and BCC lattice structures. The triplet pairing induced by the ferromagnetic spin fluctuation does not have the large eigenvalue in the \'Eliashberg equation. Moreover, our theory shows that the eigenvalue owing to the ferromagnetic spin fluctuation is suppressed by the vertex correction as the electron correlation becomes strong. Therefore, the superconducting transition temperature is too low to realize the superconductivity in the strong correlation system. This fact indicates that it is not easy for the simple ferromagnetic spin fluctuation to introduce the triplet pairing state. The result explain the fact that the superconductiviting state has not been actually found in the Ni and Pd metals, although there exist the strong ferromagnetic spin fluctuation.

On the other hand, in the very low electron density, the triplet state is also dominant owing to the ferromagnetic spin fluctuation. However, the very low density does not apply to the electron density of the realistic material. In the strong coupling system, the realistic materials usually have the intermediate or high density. The superconducting phase diagram (Fig. 3) indicates that the triplet pairing induced by the simple ferromagnetic spin fluctuation is confined to  the smaller region than that induced by the vertex correction in the intermediate density. 
Therefore, it seems that the triplet pairing is difficult to be induced by the simple ferromagnetic spin fluctuation theory such as the paramagnon theory. We think it is natural that the triplet pairing is induced by the effective interaction including the vertex correction.
\\
\\
{\bf $\diamondsuit$ All effective interaction including vertex correction}

The effect of the vertex correction is generally opposite to the effect of the RPA-like term. When the vertex correction mainly gives the advantage to the triplet pairing, the RPA-like term suppresses the triplet pairing. When the RPA-like term reflecting the antiferromagnetic (ferromagnetic) spin fluctuations induces the singlet (triplet) pairing state, the vertex correction suppresses the pairing.

When we investigate the unconventional superconductivity induced by the strong Coulomb correlation, we should consider the effective interaction as the origin of the unconventional superconductivity. In this case, it seems necessary to study not only the spin fluctuations but also the wave number dependence of the vertex correction. By analyzing the general wave number dependence of the quasi particle interaction on the basis of Fermi liquid theory, it is possible to derive generally both the singlet and triplet superconductivity.
\\   
\\
{\bf (II) \it Dimensionality and actual superconductors}

$T_{\rm c}$ in the 3-dimensional system is one-order lower than that in the 2-dimensional SQ lattice. This result indicates that the superconductivity is in favor of the 2-dimensional system such as the layered structure. The many unconventional superconductors exist in the 3-dimensional heavy fermion system. The unconventional superconductivities in the 3-dimensional system seems to exist in CeIn$_3$\cite{rf: 20} and the filled skutterudite PrOs$_4$Sb$_{12}$\cite{rf: 21} . 

CeIn$_3$ contains the SC lattice constructed by Ce atoms. CeIn$_3$ is the unconventional superconductor\cite{rf: 22} and the 3-dimensional system composed of Ce 4$f$ electrons\cite{rf: 23}. The topology of the main 3-dimensional Fermi surface\cite{rf: 23} is similar to the $d_{x^2-y^2}$-wave pairing case of Fig. 5(a) in the SC lattice. Near the half-filling in the 3-dimensional SC lattice, the antiferromagnetic order is strong near the half-filling in the 3-dimensional system. In the heavy fermion systems, the antiferromagnetic order is not so strong, and the superconductivity is also possible in the 3-dimensional SC lattice. Therefore, the superconductivity of CeIn$_3$ may be explained by the $d$-wave pairing state in SC lattice. Our result of the SC lattice has suggested that the $d_{x^2-y^2}$-wave pairing is induced, in the case that the antiferromagnetic spin fluctuation exits near $Q$=$(\pi/2,\pi/2,\pi/2)$ at Ce site. The low transition temperature $T_{\rm c}\approx 0.2$ K of CeIn$_3$ matches $T_{\rm c}$ of the 3-dimensional system lower than that of the 2-dimensional system, which is shown in our result. Actually estimating the real $T_{\rm c}$ in CeIn$_3$, we consider the effect of the Fermi energy which is reduced by the renomarization factor $z\approx 0.05$. 
 
The filled skutterudite PrOs$_4$Sb$_{12}$ has the 3-dimensional Fermi surface\cite{rf: 24} of the cubic lattice. In the study of the Sb-NQR of the PrOs$_4$Sb$_{12}$\cite{rf: 25}, the dependence of 1/$T_1$ on $T$ does not have a clear coherence peak, while it has the exponential behavior in the low temperature. The fact may indicate the possibility of the unconventional superconductivity with the full superconducting gap such as the BW state of ${}^3$He-superfluidity\cite{rf: 26}. The superconducting gap symmetry might change such as UPt$_3$\cite{rf: 27}, because the specific heat has the behavior of the double transition\cite{rf: 28}. This superconductor may be explained as the triplet pairing state owing to the effective interaction including the vertex correction. According to the present theory shown in Fig. 3, odd parity or even parity states appear mainly depending the electron density $n$ and the feature of Fermi surface.

Applying this result to the 3-dimensional Fermion system, we should reconsider the origin of the superfluidity in ${}^3$He. To make clear the origin of the superfluidity in ${}^3$He\cite{rf: 26} as the 3-dimensional system, it seems necessary to study not only the simple spin fluctuations but also the effective interaction including the wave number dependence of the vertex correction. This study is partly given in the papers$^{12,30}$.
\\
\\	
{\bf (III) \it Summary } 

Arita $et~al.$ investigated the pairing instability and the magnetic behavior in the 2- and 3-dimensional cubic lattice system on the basis of the simple spin fluctuations by the FLEX approximation.\cite{rf: 5} For the superconductivity owing to the antiferromagnetic (ferromagnetic) spin fluctuation, our result agrees with their's results. Therefore, TOPT is the reliable theory covering the theory of the spin fluctuation. 

The triplet pairing state is induced by the effective interaction including the vertex correction rather than that by the simple ferromagnetic fluctuation. The effective interaction including the vertex correction plays the important role for the triplet superconductivity in the 3-dimensional system as well as the 2-dimensional system$^{11,12)}$. When we investigate the unconventional superconductivity in the strong correlated system, the mechanism should be considered  owing to not only the simple spin fluctuation but also the effective interaction including the vertex correction. A part of this study is given in the papers\cite{rf: 29}.

\section{Acknowledgment}
H. F. is also grateful to Prof. H. Harima, Dr. H. Kotegawa and S. Kawasaki. In this study, H. F. thanks the Yukawa Institute Facility and the Supercomputer Center, Institute for Solid State Physics, University of Tokyo. Our work is partly supported by the Ministry of Education, Science, Sports and Culture.

\clearpage
\begin{figure}
\epsfile{file=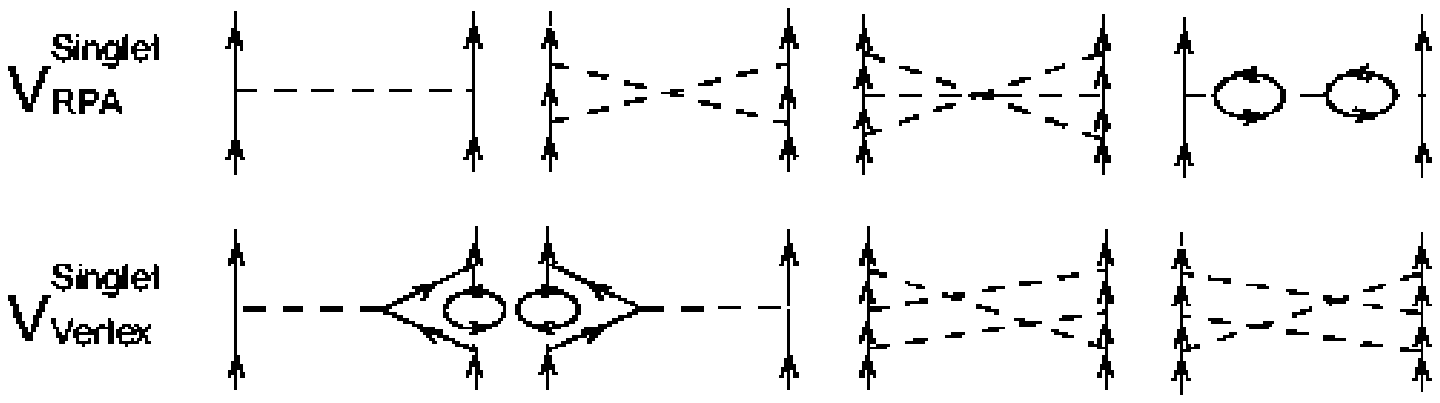,height=4cm}
\caption{Diagrams for the effective interaction of singlet pairing within the third-order perturbation with respect to $U$. The solid line is the bare Green's function $G_0$. The broken line is the Coulomb interaction $U$. The broken line of $U$ connects only solid lines possessing opposite spins. The two external lines have the opposite spins. The effective interaction is divided into the RPA-like part and the vertex correction. The latter begins with the third-order terms.}
\label{fig:1}
\end{figure}
\begin{figure}
\epsfile{file=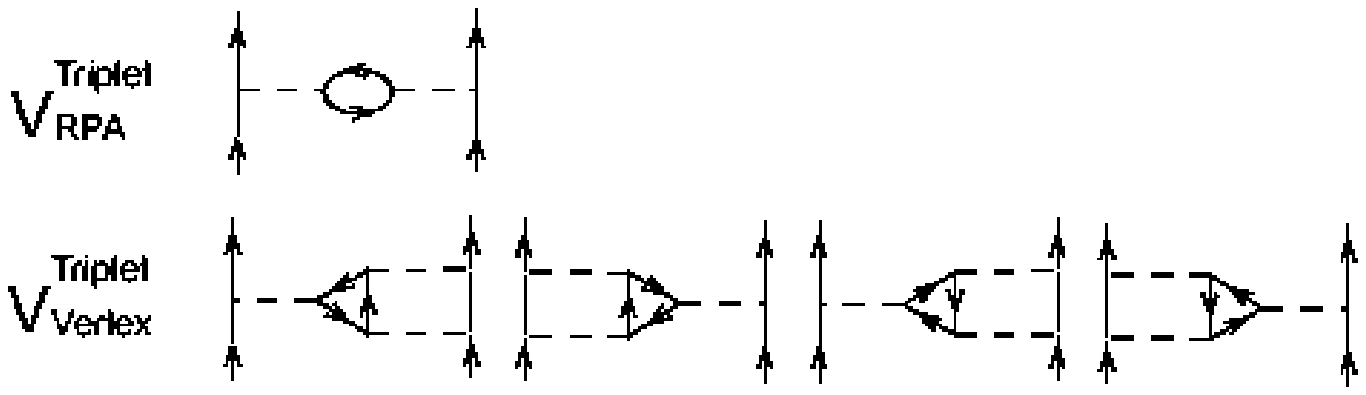,height=4cm}
\caption{Diagrams for the effective interaction of the triplet pairing within the third-order perturbation. The two external lines have the parallel spins. The RPA-like part and the vertex correction of the pairing interaction are given by only the second-order and the third-order terms, respectively.}
\label{fig:2}
\end{figure}
\begin{figure}
\epsfile{file=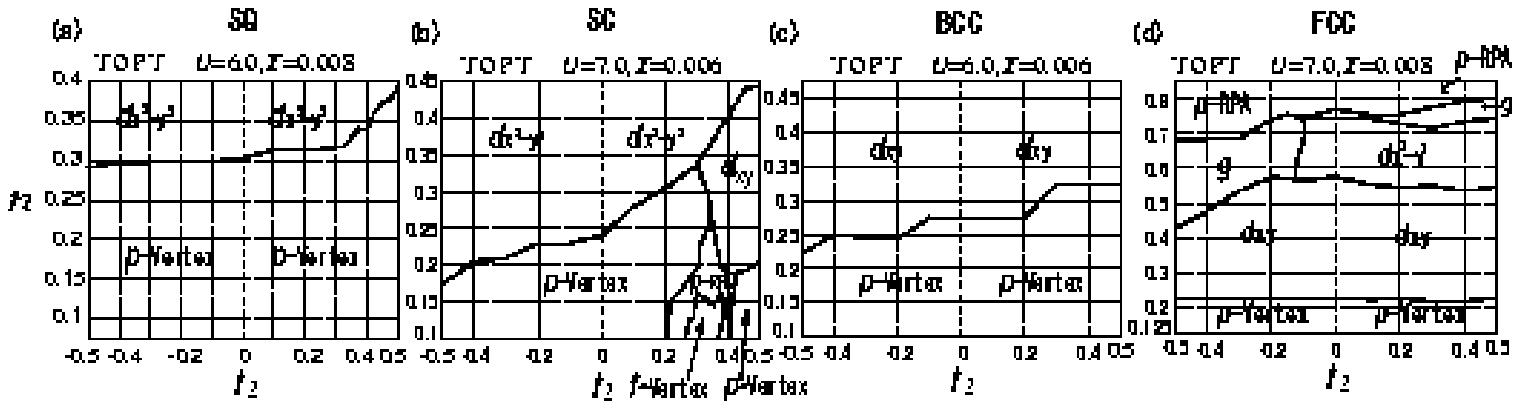,height=4.5cm}
\caption{The superconducting phase diagrams for the dominant pairing symmetry. $t_2$ and $n$ are the next-nearest hopping integral and the density, respectively. The half-filling density corresponds to $n$=0.5.}
\label{fig:3}
\end{figure}

\begin{figure}
\epsfile{file=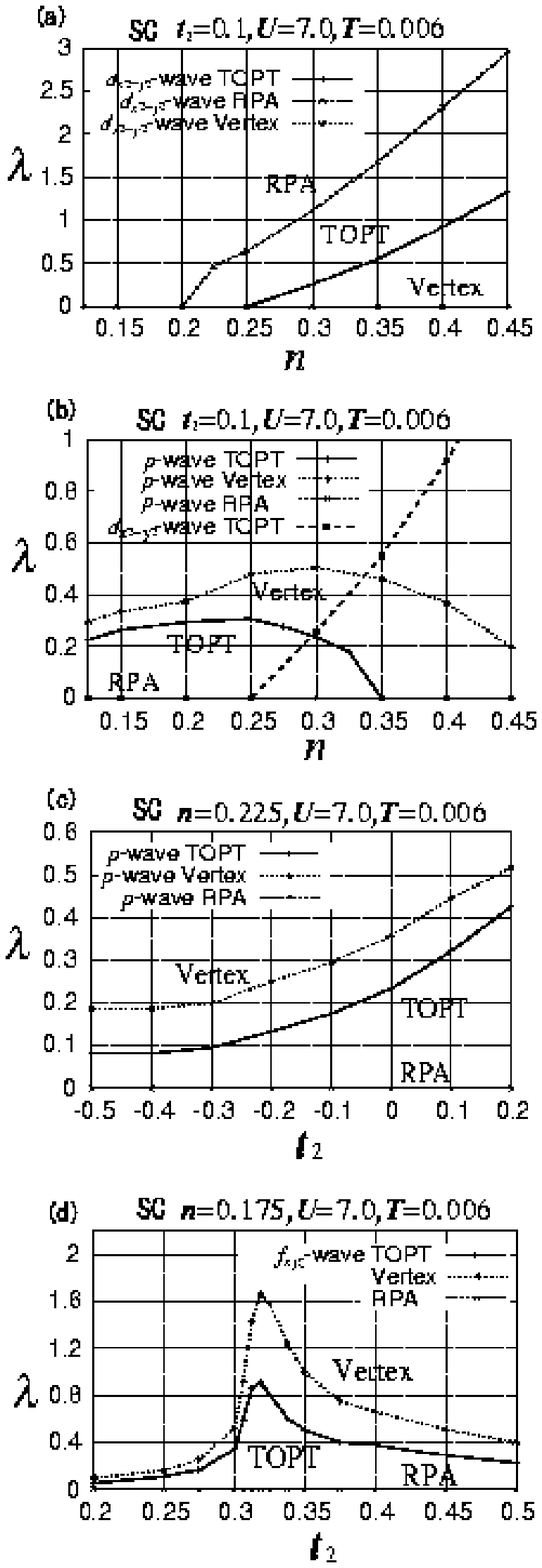,height=22.0cm}
\caption{The $n$- and $t_2$-dependence of the eigenvalue $\lambda$ obtained by TOPT, vertex and RPA-like terms for the $p$-, $d_{x^2-y^2}$- and $f_{xyz}$-wave pairing states in the SC lattice structure.}
\label{fig:4}
\end{figure}

\begin{figure}
\epsfile{file=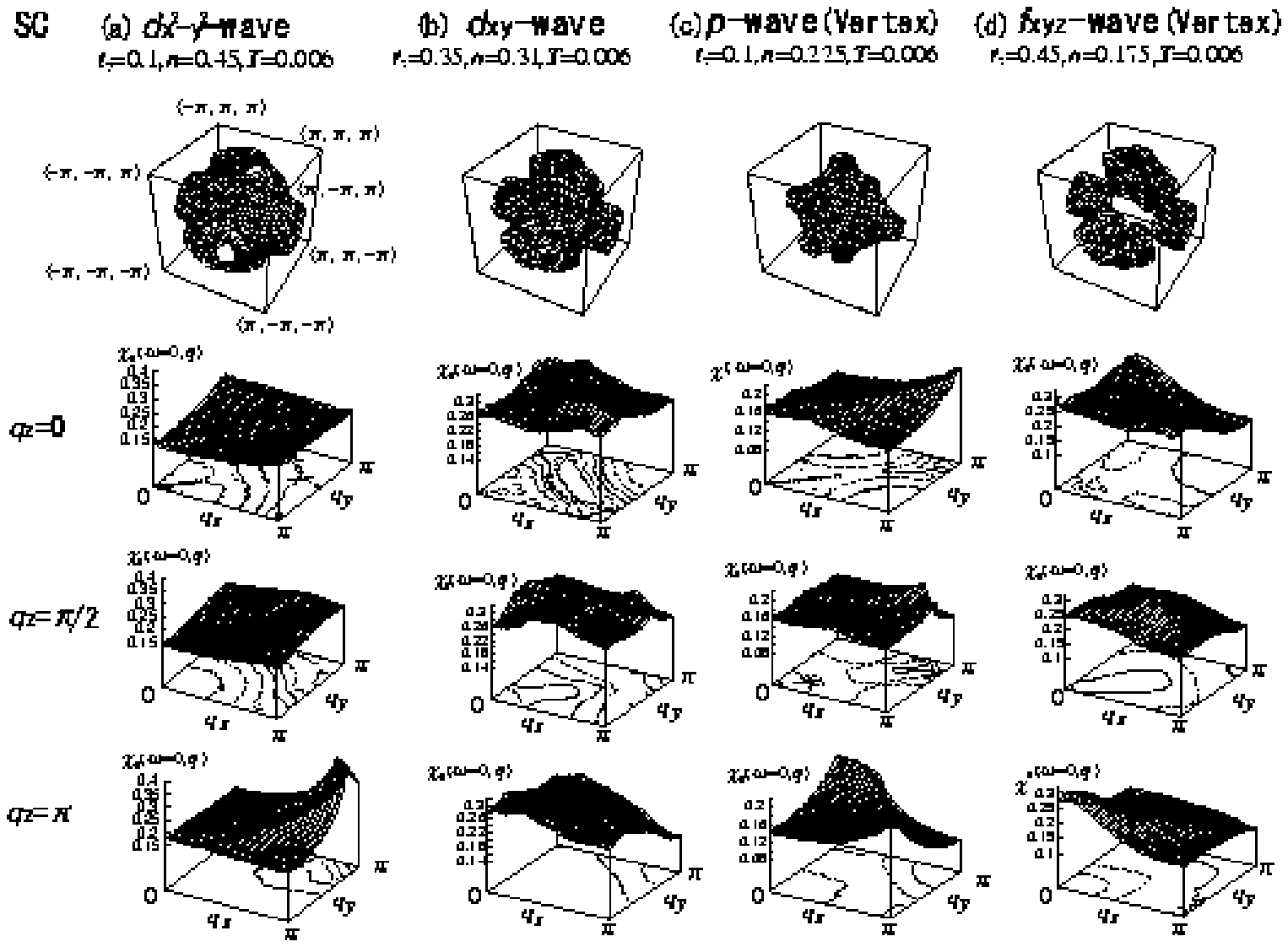,height=12cm}
\caption{The bare susceptibility $\chi_0(q, \omega_n=0)$ and Fermi surface in a quarter first-Brillouin-zone of the SC lattice structure. (a) and (b) are in the case that the $d$-wave pairing states are induced by the spin fluctuation. (c) and (d) are in the case that $p$- and $f$-wave pairing states are dominant owing to the Vertex correction.}
\label{fig:5}
\end{figure}

\begin{figure}
\epsfile{file=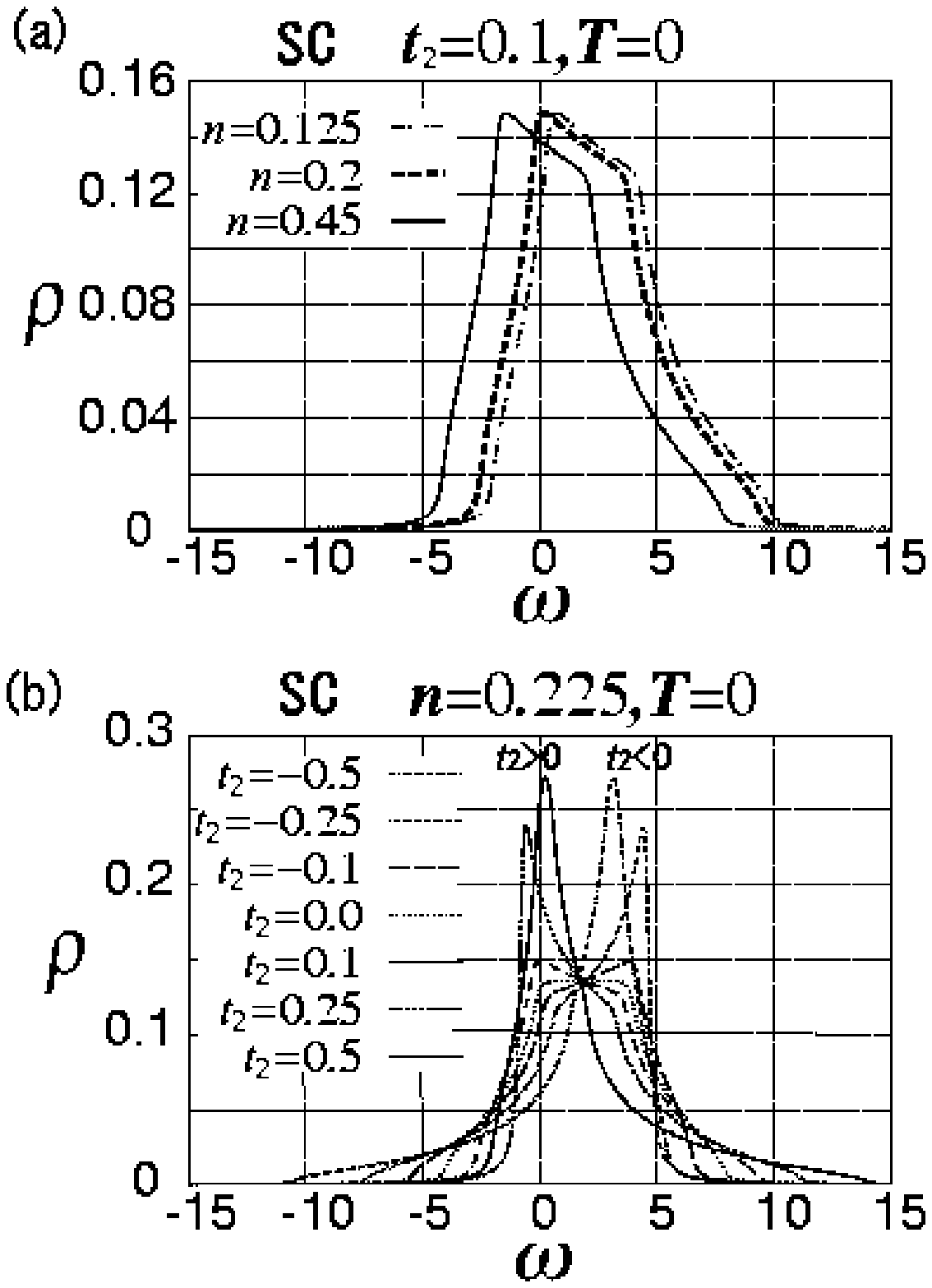,height=12.0cm}
\caption{The $n$- and $t_2$-dependence of the density of states at $T$=0 and $U$=0 for the SC lattice structure. The parameters are fixed $t_2$=0.1 for (a) and $n$=0.225 for (b), respectively.}
\label{fig:6}
\end{figure}

\begin{figure}
\epsfile{file=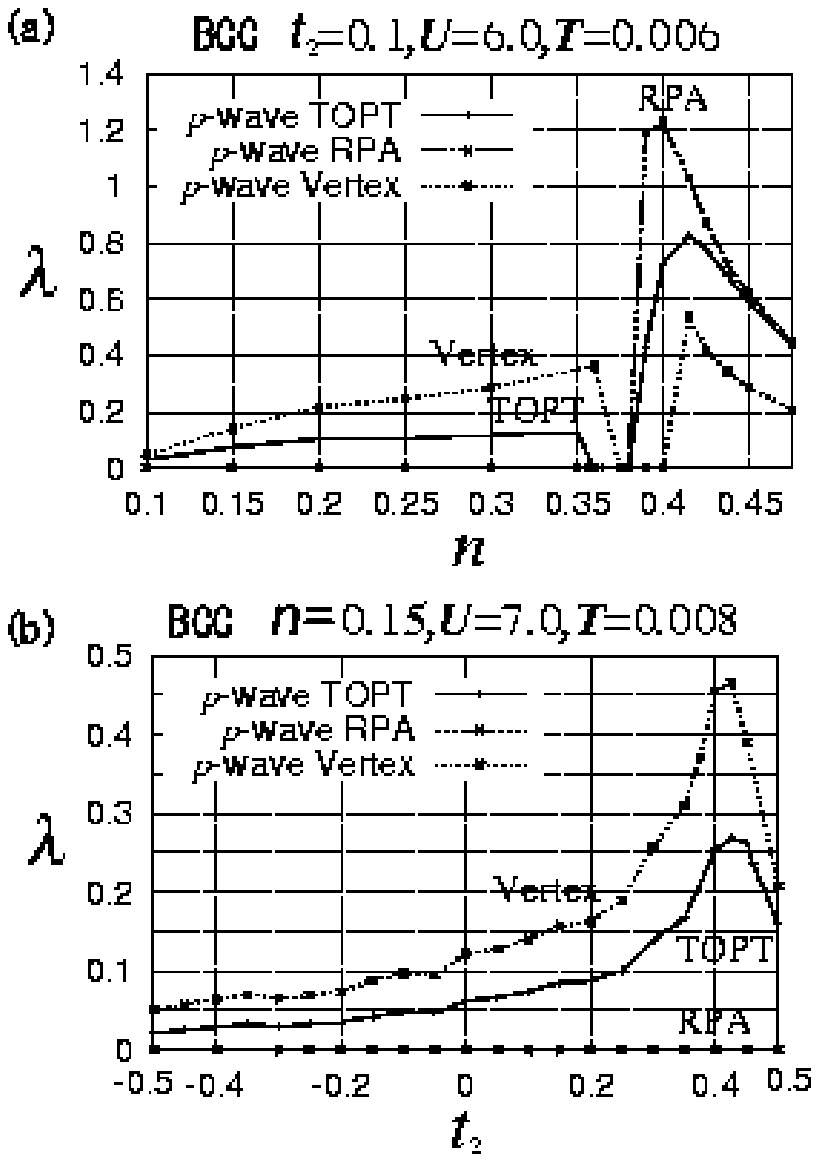,height=12.0cm}
\caption{(a) The $n$-dependence of the eigenvalue $\lambda$ for the $p$- and $d_{x^2-y^2}$-wave pairing states for the BCC lattice structure. The parameters of (a) are fixed at $t_2$=0.1, $U$=6.0 and $T$=0.006. (b) The $t_2$-dependence of the eigenvalue $\lambda$ for the $p$- and $d_{x^2-y^2}$-wave pairing states for the BCC lattice structure. The parameters are fixed at $n$=0.15 and $T$=0.008.}
\label{fig:7}
\end{figure}

\begin{figure}
\epsfile{file=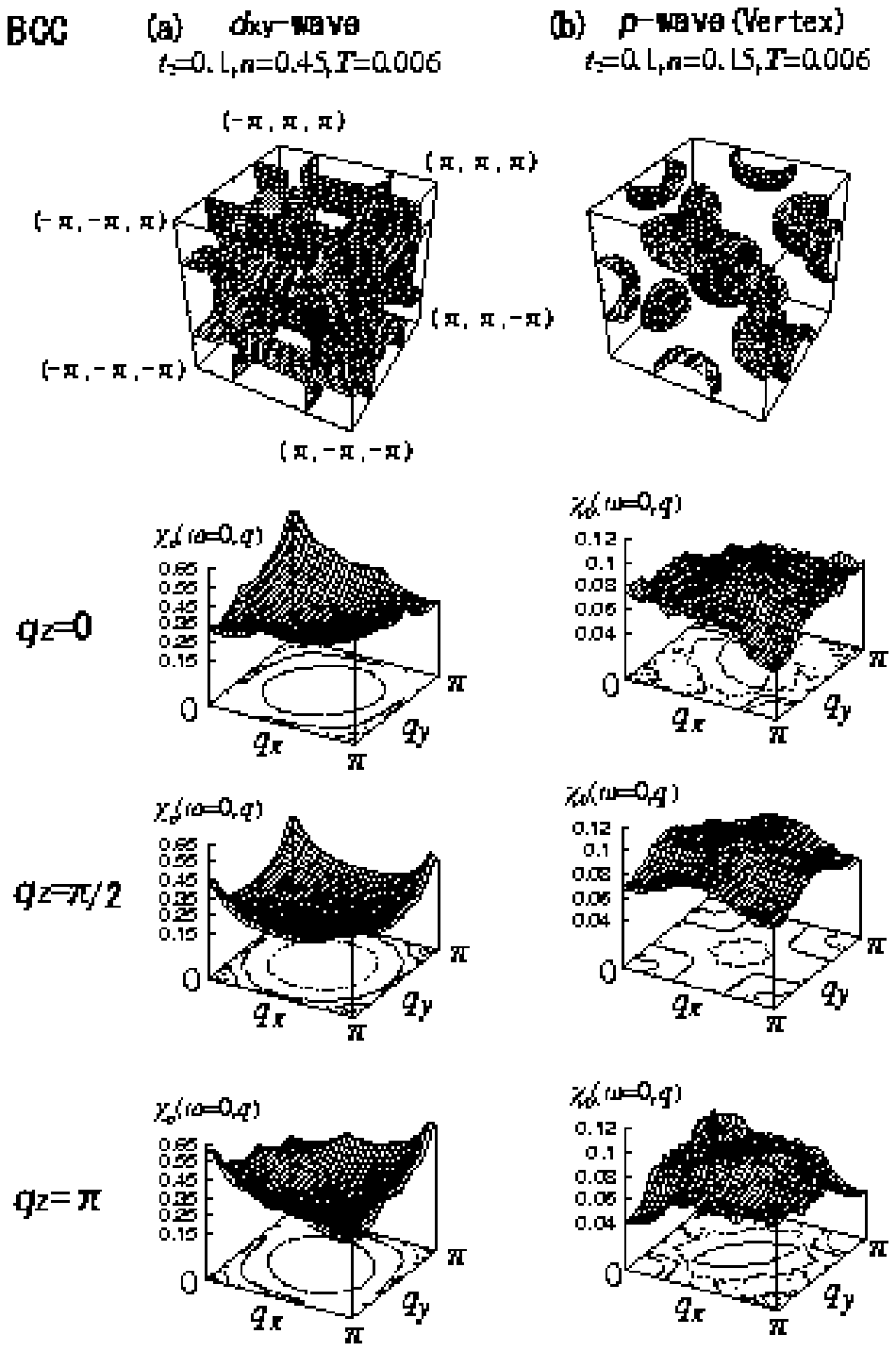,height=13.0cm}
\caption{The bare susceptibility $\chi_0(q, \omega_n=0)$ and Fermi surface for the BCC lattice structure for the BCC lattice structure. (a) is in the case that the $d_{xy}$-pairing state is induced by the spin fluctuation. (b) is in the case that the $p$-wave pairing is encouraged by the Vertex correction. }
\label{fig:8}
\end{figure}

\begin{figure}
\epsfile{file=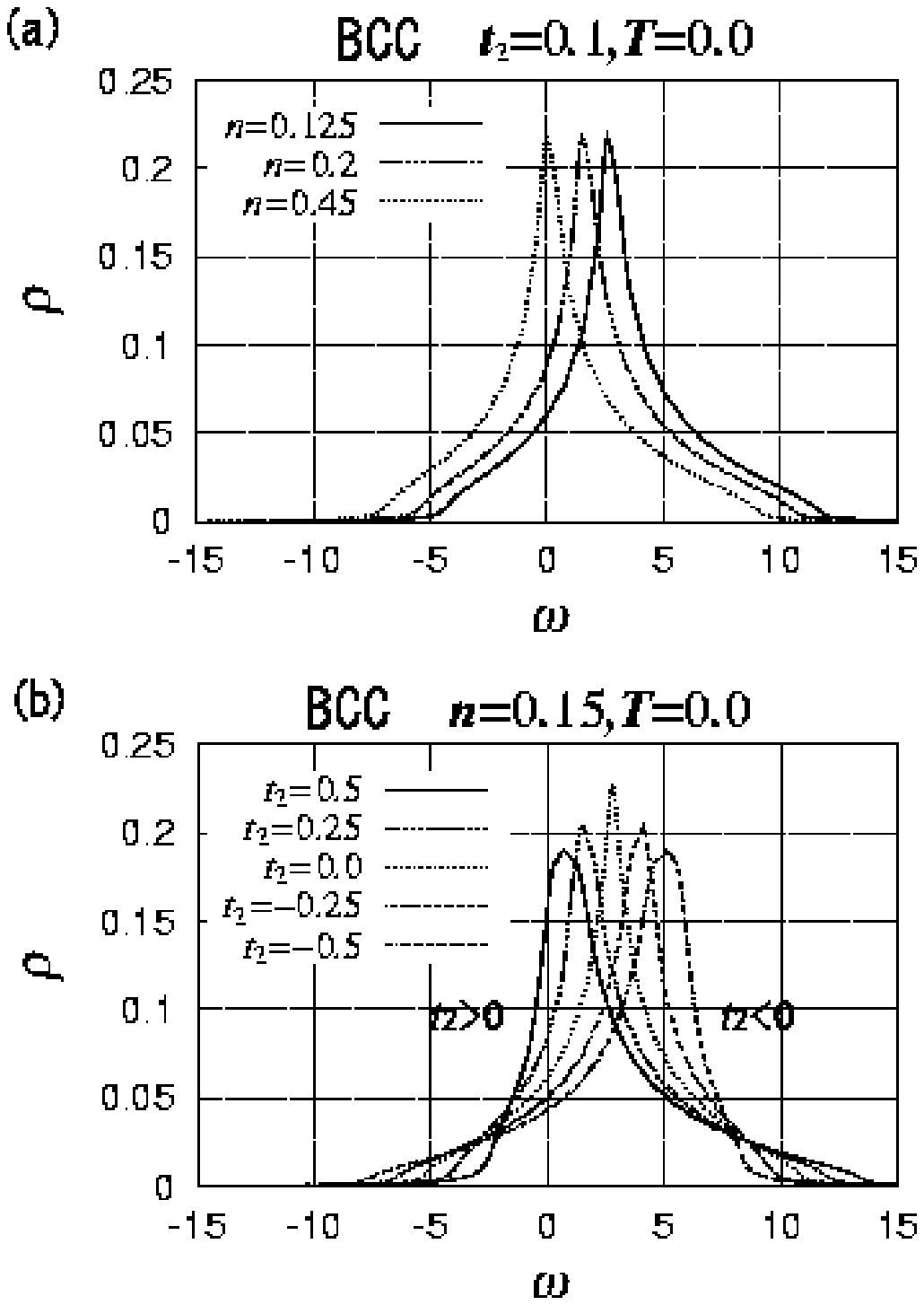,height=12.0cm}
\caption{The $n$- and $t_2$-dependence of the density state at $T$=0 for the BCC lattice structure. The parameters are fixed $n$=0.175 and $t_2$=0.1, respectively.}
\label{fig:9}
\end{figure}

\begin{figure}
\epsfile{file=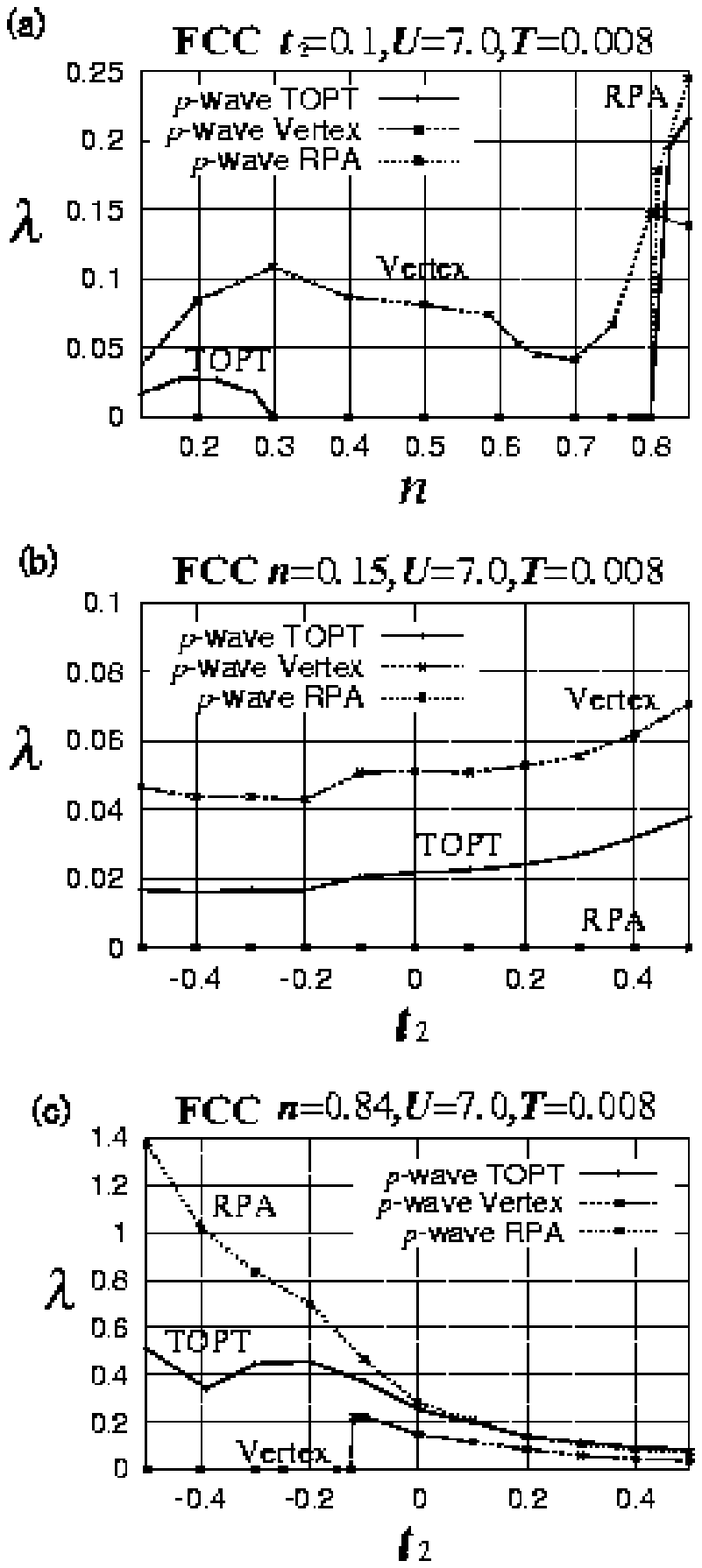,height=20.0cm}  
\caption{(a) The $n$-dependence of the eigenvalue $\lambda$ for the $p$-wave pairing state for the FCC lattice structure. The parameter is fixed at $t_2$=0.1. (b), (c) The $t_2$-dependence of the eigenvalue $\lambda$ for the $p$-wave pairing state for the FCC lattice structure. The parameters are fixed at $n$=0.15 and 0.84, respectively.}
\label{fig:10}
\end{figure}

\begin{figure}
\epsfile{file=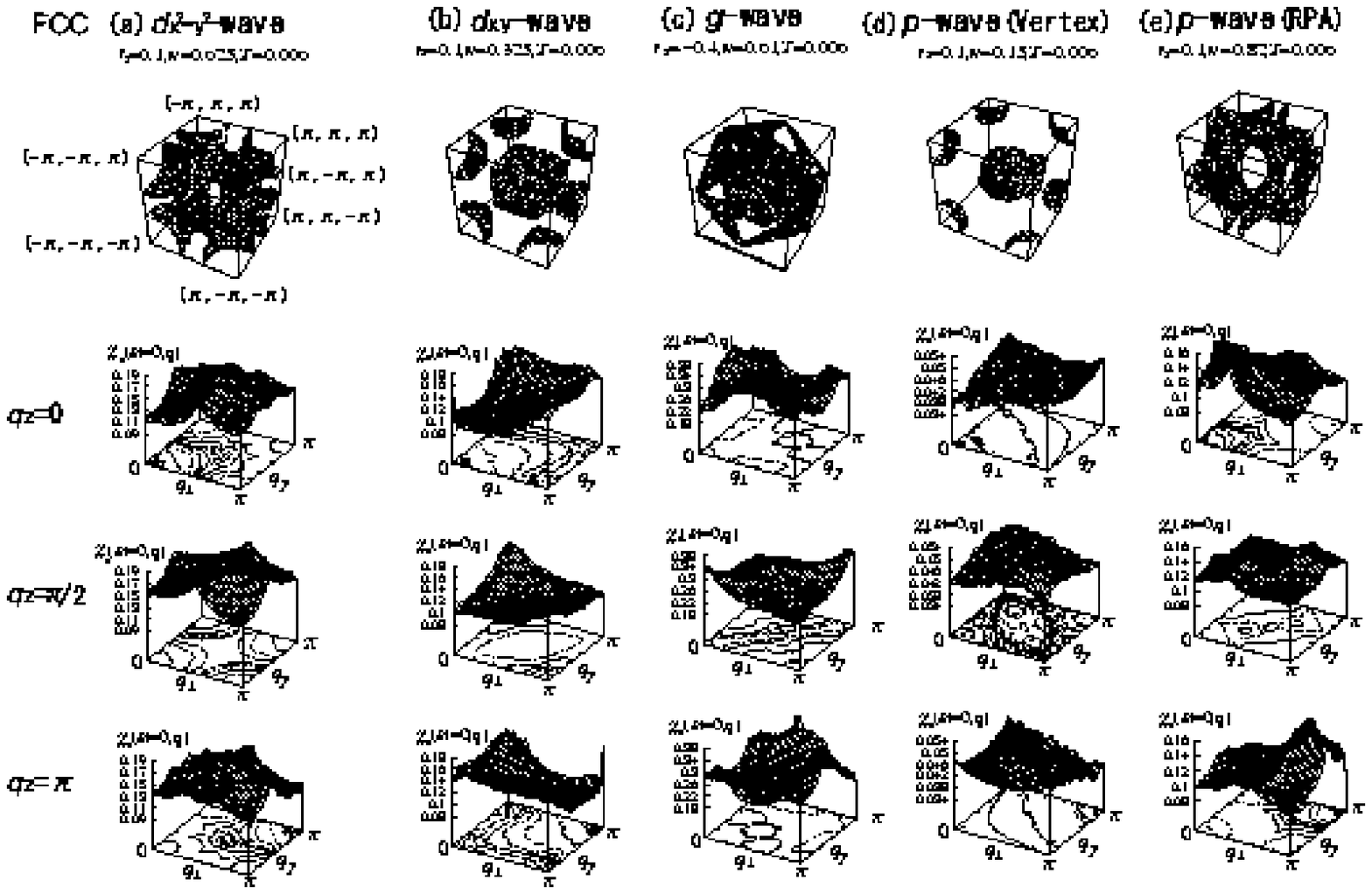,height=10.5cm}
\caption{The bare susceptibility $\chi_0(q, \omega_n=0)$ and Fermi surface for the FCC lattice structure. (a), (b) and (c) exhibit $\chi_0(q, \omega_n=0)$ in the cases of the $d$- and $g$-wave pairing states due to the RPA-like term. (d) corresponds to the $p$-wave pairing which is encouraged by the Vertex correction term. (e) corresponds to the $p$-wave pairing owing to the RPA-like term.}
\label{fig:11}
\end{figure}

\begin{figure}
\epsfile{file=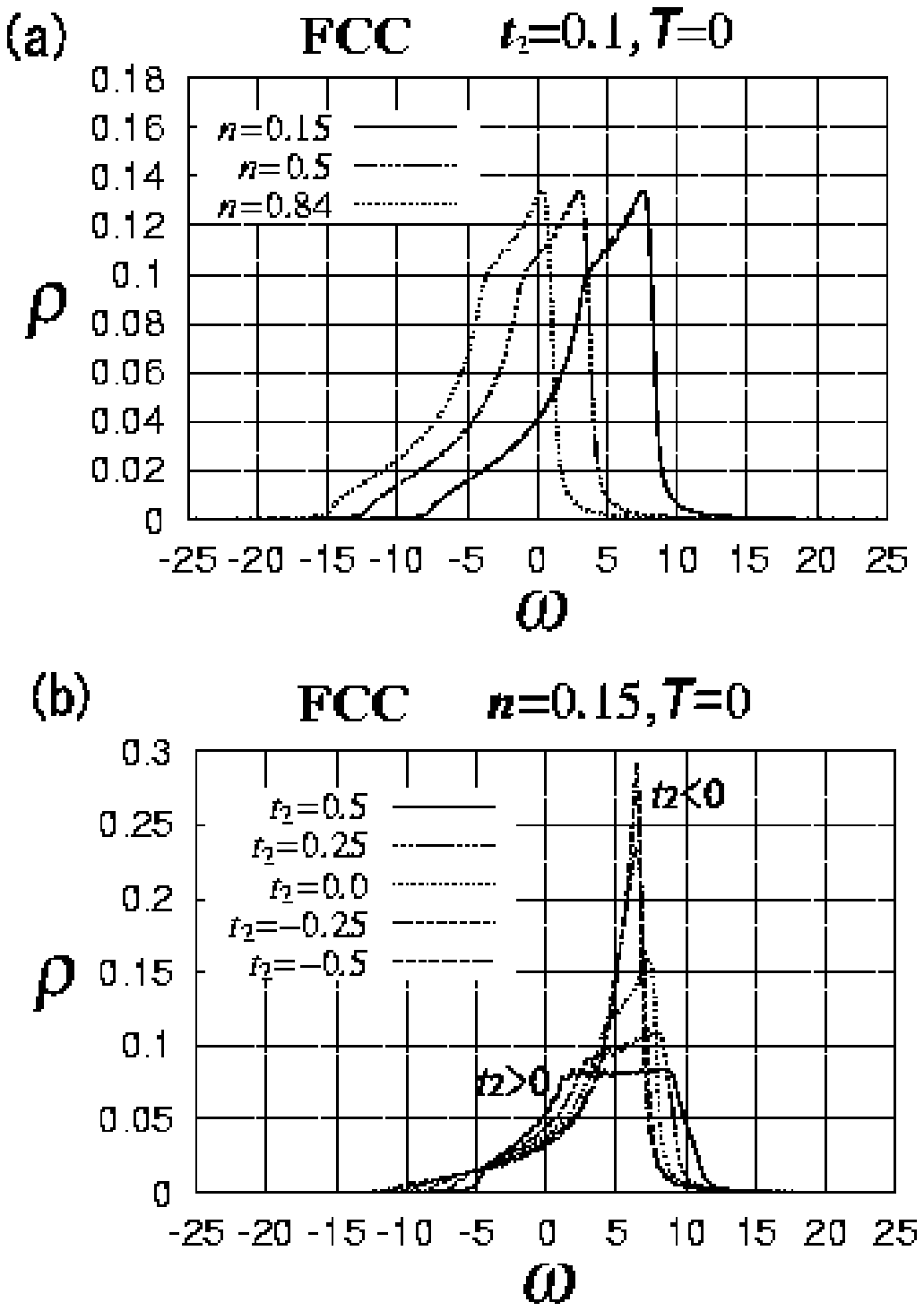,height=12.0cm}
\caption{The $n$- and $t_2$-dependence of the density of states at $T$=0 for the FCC lattice structure. The parameters are fixed $t_2$=0.1 and $n$=0.15, respectively.}
\label{fig:12}
\end{figure}

\begin{figure}
\epsfile{file=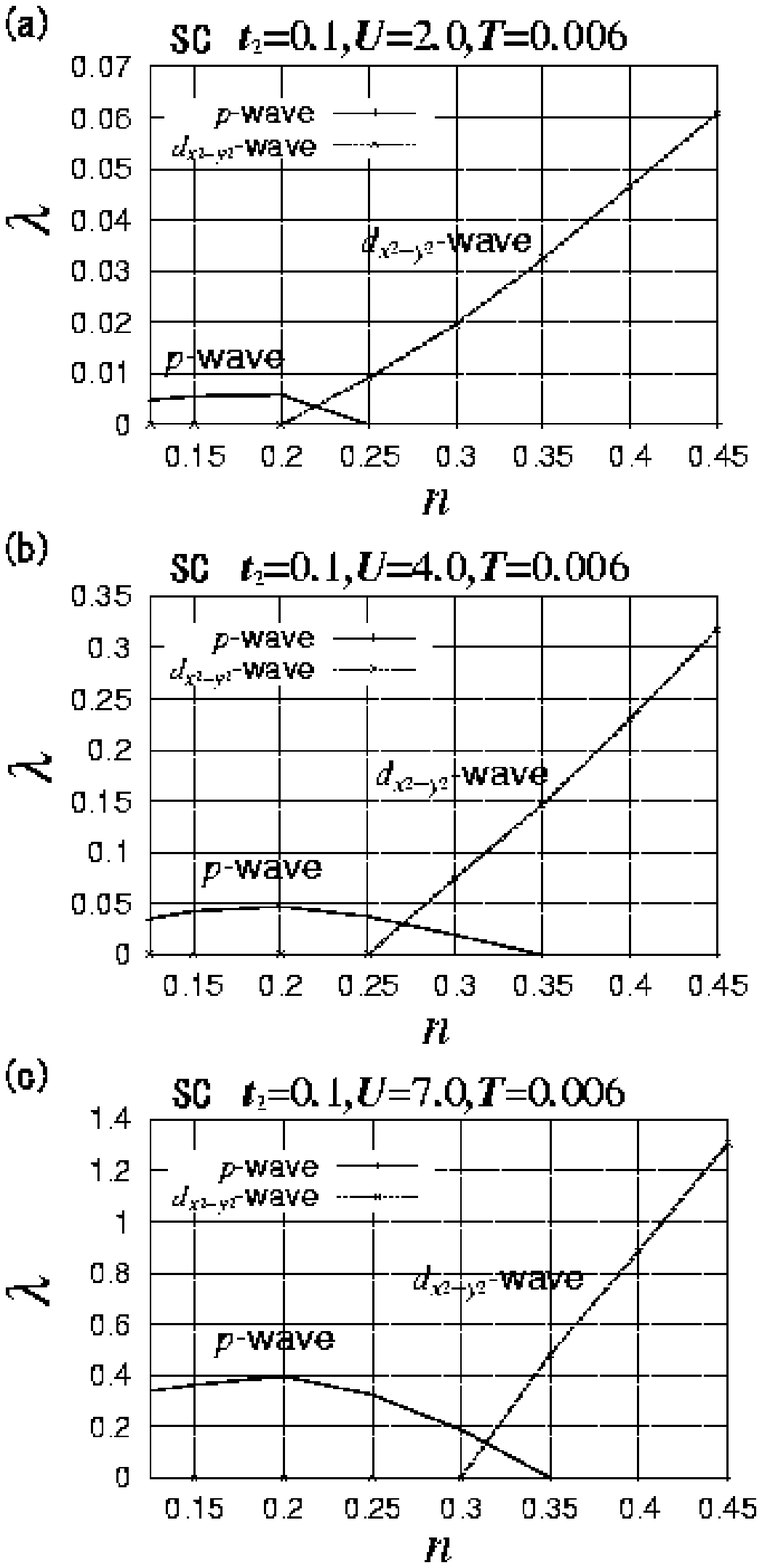,height=20cm}
\caption{The $U$-dependence of the eigenvalue $\lambda$ obtained by TOPT, vertex and RPA-like terms for the $p$-, $d_{x^2-y^2}$-wave pairing states in the SC lattice structure.}
\label{fig:13}
\end{figure}

\begin{figure}
\epsfile{file=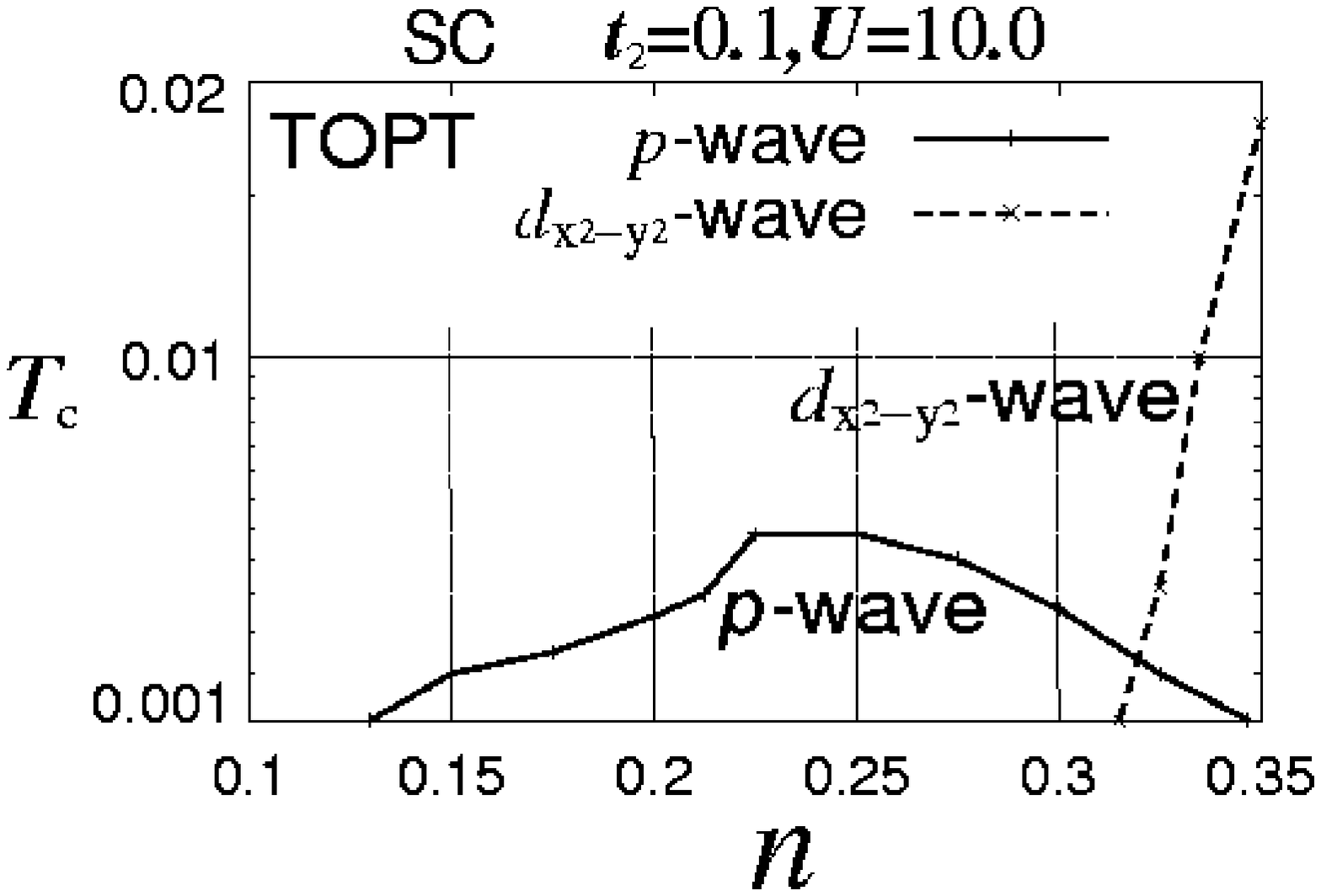,height=7cm}
\caption{The $n$-dependence of the superconducting transition temperature $T_{\rm c}$ for the $p$- and $d_{x^2-y^2}$-wave pairing states for the SC lattice structure. $T_{\rm c}$ is calculated by TOPT. The parameters are $t_2$=0.1, $U$=10.0. The unit of energy is the hopping transfer $t_1$.}
\label{fig:14}
\end{figure}
\end{document}